\documentclass[12pt, a4paper]{article}
\usepackage{pstricks}

\usepackage{tikz}
\usetikzlibrary{matrix,arrows,decorations.pathmorphing}

\usetikzlibrary{shapes,arrows}
\usepackage{pgf}
\usepgfmodule{shapes}
\usepgfmodule{plot}
\usetikzlibrary{decorations}
\usetikzlibrary{arrows}
\usetikzlibrary{snakes}

\tikzstyle{decision} = [diamond, draw,
    text width=4.5em, text badly centered, node distance=3cm, inner sep=0pt]
\tikzstyle{block} = [rectangle, draw,
    text width=5em, text centered, rounded corners, minimum height=4em]
\tikzstyle{line} = [draw, -latex']
\tikzstyle{cloud} = [draw, ellipse]

\usepackage{latexsym}
\usepackage{amsmath, amscd}
\usepackage{amssymb}
\usepackage{graphicx}
\usepackage{euscript}
\usepackage{amsfonts}
\usepackage{verbatim}
\usepackage{epsfig}
\usepackage{hyperref}
\usepackage[margin=2cm,bottom=2.0cm]{geometry}

\begin{document}

\begin{titlepage}
\titlepage
\begin{flushright}{CERN-PH-TH/2012-051, TUW-12-04}
\end{flushright}
\vskip 2.5cm
\begin{center} \bf \huge Algebraic description of G-flux in F-theory:\\ new techniques for F-theory phenomenology
\end{center}
\vskip 2.3cm

\begin{center}
{\bf \large Andreas P. Braun$^1$, Andr\'es Collinucci$^{2, 3}$ and Roberto Valandro$^4$}
\vskip 0.3cm
\em 
{\it$^1$ Institute for Theoretical Physics,
Vienna University of Technology, Austria }
\\[.5cm]
{\it $^2$ Theory Group, Physics Department, CERN CH-1211 Geneva 23, Switzerland}
\\[.5cm]
{\it $^3$ Physique Th\'eorique et Math\'ematique
Universit\'e Libre de Bruxelles, C.P. 231, 1050 Bruxelles, Belgium}
\\[.5cm]
{\it $^4$ II. Institut fuer Theoretische Physik, Hamburg University, Germany}

\vskip 2.5cm

\large \bf Abstract
\end{center}

We give a global algebraic description of the four-form flux in F-theory. We present how to compute its D3-tadpole and how to calculate the number 
of four-dimensional chiral states at the intersection of 7-branes directly in F-theory.  We check that, in the weak coupling limit, we obtain the 
same results as using perturbative type IIB string theory. We develop these techniques in full generality. However, they can be readily applied 
to concrete models, as we show in an explicit example.

\vskip2cm

\vskip1.5\baselineskip

\vfill
\vskip 2.mm
\noindent {\small \it Based on a talk by R.~V.~at the XVII European Workshop on String Theory 2011, Padova, Italy, 5-9 September 2011.
\noindent }
\end{titlepage}

\tableofcontents

\section{Introduction}



The two basic building blocks for F-theory model building are an elliptically fibred Calabi-Yau fourfold $X_4$, and a four-form $G_4$ living on it. 
The first is important to understand which gauge group and which matter representations we can realize.
The second is responsible for moduli stabilization and for generating four-dimensional chiral modes.
In fact, the Calabi-Yau fourfold $X_4$ encodes both the type IIB compactification manifold $B_3$ and the configuration of the 7-branes on which gauge theory can live. 
The manifold $X_4$ is an elliptic fibration over the base $B_3$. The complex structure of the elliptic fibre is identified with the type IIB axion-dilaton $\tau=C_0+i\,e^{-\phi}$. 
The 7-brane is located at the locus over which the elliptic fibre degenerates, corresponding to $\tau\rightarrow i\infty$ (up to $SL(2,Z)$). 
The requirement to have ${\cal N}=1$ supersymmetry in four dimensions impose the manifold $X_4$ to be a Calabi-Yau fourfold.
 In type IIB, which we are discussing here, this comes from the charge cancelation of the 7-branes, or equivalently from the backreaction of the branes (= deficit angles) on the
geometry.
The second important object, i.e. $G_4$, encodes the type IIB bulk three-form field strengths $H_3$ and $F_3$ and the gauge field strength $F_2$ living on the branes. 
One notes that $G_4$ encodes fluxes relevant for both moduli stabilisation and chirality. 

From the perspective of type IIB string theory, the two-torus is auxiliary. However, it becomes a `real' part of the compactification in M-theory. In fact, one can effectively 
realize the F-theory idea through a chain of dualities (see \cite{Denef:2008wq} for a review):
 \begin{equation}\begin{array}{ccccc}
 \mbox{IIB on }S^1_B \times B_3&  &  \mbox{IIA on } S^1_A \times B_3 & & \mbox{M-th on } X_4= T^2 \tilde{\times} B_3. \\ 
 \mbox{\scriptsize $R_B\rightarrow \infty$}   && \mbox{\scriptsize $R_A\rightarrow 0$} && \mbox{\scriptsize Size$(T^2)\rightarrow 0$} \\
 &\xrightarrow[]{T-duality}&   &\xrightarrow[]{Uplift} &\\
 {\cal H}_3,{\cal F}_3,F_2   && {\cal H}_3,{\cal F}_4,F_2 && G_4 \\
 \mbox{\footnotesize along $B_3$}   &&  && \mbox{\footnotesize one leg along fiber} \\
 \end{array}\nonumber\end{equation}
In order to have type IIB four dimensional compactifications, one has to take the fibre size to zero on the M-theory side. This is called the F-theory limit. 
In order for the type IIB background flux to preserve the four dimensional Poincar\'e invariance, the four-form $G_4$ must have one and only one leg along the elliptic fibre $T^2$.

To capture all possible consistency constraints, we need a global construction. 
The $CY_4$ is the better understood one (for recent GUT-like global models see \cite{Collinucci:2008zs, Marsano:2009ym, Collinucci:2009,Blumenhagen:2009up, Blumenhagen:2009yv, Marsano:2009wr, Grimm:2009yu, Chen:2010ts}).
Only in the past year has a deep study of global constructions of the $G_4$ flux been undertaken \cite{Grimm:2011tb,Braun:2011zm,Marsano:2010ix,Krause:2011xj, Krause:2012yh}. 

In \cite{Braun:2011zm} we constructed a global $G_4$ flux in an algebraic way. This construction is very explicit and easy to use for model building.
We identified this flux with the $U(1)$ flux on a specific $I_1$-brane. This flux generates chiral modes for matter living on the intersection of this branes with the GUT brane.

This note summarizes the main results of \cite{Braun:2011zm}, to which we refer more details.

\section{$G_4$-flux on a smooth $CY_4$}

\subsection{New algebraic cycles}

We want to construct a supersymmetric (i.e. of type $(2,2)$ and primitive), properly quantized ($G_4+\frac{c_2}{2}\in H^4(\mathbf{Z})$) and Poincar\'e invariant (only one leg along fiber) four-form flux.

Our strategy is to describe $G_4$ in terms of algebraic integral cycles via Poinca\'re duality.
\begin{equation}\label{gammainstuff}
 \gamma_4 \in H_{2,2}(CY_4)\cap H_4(CY_4,\mathbf{Z}) 
 \end{equation}
s.t. $\int_{D_i\cdot D_j}G_4 = \int_{D_i\cdot B_3}G_4=0$,
with $D_i$ elliptic fibrations over divisors of $B_3$.

Let us consider a smooth Calabi-Yau four-fold $X_4$ described by the Weierstrass equation
\begin{equation}\label{Weq}
W\,: \,\,\,y^2=x^3 + f\, x \,z^4 +g \,z^6\:
\end{equation}
in the ambient space $X_5=\mathbf{P}_{2,3,1}({\cal O}_{B_3}\oplus {\cal O}_{B_3}\oplus K_{B_3})$, where $f$ and $g$ are sections of $K_{B_3}^{-4}$ and $K_{B_3}^{-6}$, respectively, 
and $K_{B_3}$ is the canonical bundle of $B_3$. For simplicity and without affecting the results, we will work in the patch $z=1$ in the following.

The 7-brane locations are given by the discriminant locus
\begin{equation}
 \Delta = f^3+g^2 = 0 \, .
\end{equation}
For generic $f$ and $g$ there is only one single $I_1$-brane.
In the weak coupling limit, this corresponds to one O7-plane and one invariant D7-brane \cite{Sen:1997gv}.
We are interested in identifying the $G_4$ flux corresponding to a supersymmetric $F_2$ flux on this brane.

\

For generic $f$ and $g$, all algebraic four-cycles are given by intersecting two equations with $W$, i.e. from the point of view of the Poincar\'e dual four-forms
$H^{2,2}\cap H^4(\mathbf{Z}) = H^{1,1}(\mathbf{Z}) \wedge H^{1,1}(\mathbf{Z})$.
But $G_4$ $\in$ $H^{1,1}\wedge H^{1,1}$ violates the Poincar\'e invariance.
For smooth fourfolds, such four-forms always have either two or no legs along the elliptic fibre. Hence, it would seem a priori impossible to create acceptable holomorphic four-cycles by algebraic means.

The main result of our work \cite{Braun:2011zm} is an solution to this puzzle. Before looking for such cycles algebraically, the fourfold must first assume the appropriate configuration.
The Weierstrass equation takes an interesting form if we restrict the complex structure of the Calabi-Yau fourfold $X_4$ such that $g$ takes the form: 
\begin{equation}
g\equiv \psi^2-\rho\tau \qquad \rightarrow \qquad 
W\,:\,\,\,(y-\psi)(y+\psi) + \rho\,\tau = x \, ( x^2 + f ) \:.
\end{equation}
The important point is that now new algebraic four-cycles of $X_4$ arise. These are four-cycles of $X_5$ that lie in $X_4$:
\begin{equation}
 \gamma_4:\qquad y = \psi \qquad\cap\qquad x = 0 \qquad\cap\qquad \rho=0 \qquad\qquad \subset X_5\:.\nonumber
\end{equation}
Hence there are now new elements of $H^{2,2} \cap H^4(\mathbf{Z})$ that are \emph{not} intersections of two divisors in $X_4$.

Using this, we are able to define a four-form flux that is Poincar\'e invariant:
\begin{equation}\label{G4sm}
 G_4 = \gamma_4^{(PD_{CY_4})} - s 
\end{equation}
with $s$ a properly chosen element of $H^{1,1}(\mathbf{Z})\wedge H^{1,1}(\mathbf{Z})$ (see \cite{Braun:2011zm} for details).

\

The algebraic definition of $G_4$ allows for a direct computation of its induced D3-charge in terms of intersections on the base:
\begin{equation}
 Q_{D3} = \tfrac12 \int_{CY_4} G_4\wedge G_4 = \tfrac12 \left( \gamma_4\cdot \gamma_4|_{CY_4} - \int_{\gamma_4} s\right) = -\int_{B_3} c_1(B_3) \cdot [\rho] \cdot[\tau] \, ,
\end{equation}
where $[\rho]$ and $[\tau]$ denote the divisor classes of the corresponding polynomials.


We have thus shown how to construct a certain class of $G_4$-fluxes very straightforwardly by manipulating the Weierstrass model. The price, however, is loss of intuition. In the next section, we will make contact with perturbative IIB through a more elaborate version of this construction, whereby the interpretation of the flux will be that of a DBI worldvolume flux on a D7-brane.

\subsection{Weak coupling limit}

There exists an analogous construction in the weak coupling limit \cite{Sen:1997gv}. Here, $f$ and $g$ are parametrized as 
\begin{equation}
 f=\epsilon\,\eta-3h^2 \qquad g = h(\epsilon \,\eta-2h^2) + \chi \:.
\end{equation}
The weak coupling limit corresponds to taking the parameter $\epsilon$ to zero ($\epsilon\rightarrow 0$). After the limit, the $I_1$-brane splits into one O7-plane and one D7-brane wraping
\begin{equation}
 O7: \,\,h=0 \qquad \qquad D7:\,\, \eta^2 - \xi^2 \chi = 0  \qquad \mbox{with}\qquad \xi^2 = \,h \:. 
\end{equation}
The last equation describes the type IIB Calabi-Yau threefold as a double cover of the base manifold $B_3$. The orientifold involution is $\xi\mapsto -\xi$.

In this setup, new algebraic cycles arise when the complex structure of the fourfold is restricted to
\begin{equation}
\chi = \psi^2-\rho\tau \, .
\end{equation}
On the type IIB side, this corresponds to fixing some of the D7-brane moduli. The corresponding flux is given by the Poincar\'e dual of the two-cycle given by the {\it non-transverse} intersection of two divisors:
\begin{equation}
 F_2 \,\,\,\leftrightarrow\,\,\, \eta =  \xi \psi \qquad\cap\qquad \rho = 0 \qquad \qquad\subset D7 \, .
\end{equation}
This is in complete analogy to our result: some moduli in the Weierstrass equation are fixed, which allows for a flux Poincar\'e dual to the four-cycle given by the {\it non-transverse} intersection of three divisors:
\begin{equation}
 G_4 \,\,\leftrightarrow\,\, y = \psi \qquad\cap\qquad x=h \qquad\cap\qquad \rho = 0 \qquad \subset \, CY_4 
\end{equation}

We claim that  $G_4$ corresponds to the $F_2$ flux on the D7-brane. This is supported by two facts: 1) They stabilize the same moduli ($\chi=\psi^2-\rho\tau$) and 2) they have the same D3-charge
\begin{equation}
-\tfrac12 \int_{D7} F_2\wedge F_2 = \tfrac12 \int_{CY_4} G_4\wedge G_4 \:.
\end{equation}

\section{Singularities and $G_4$ induced chirality}

\subsection{D7/Image-D7 configuration}

The smooth Calabi-Yau fourfold describes the situation with only one recombined 7-brane (in the weak coupling limit this splits into an orientifold plane and one invariant D7-brane). This configuration cannot produce chiral modes. 
This means that in order to have such modes, we need a singularity in the fourfold.
In fact, massless matter in F-Theory arises from singularities at codimension two, in the base. We now consider the simplest case, which is given by a conifold singularity along a curve in the base $B_3$. 
To realise this, one takes $g\equiv \psi^2$. Correspondingly the Weierstrass equation \eqref{Weq} becomes
\begin{equation}
 (y-\psi)(y+\psi) = x (x^2+f)
\end{equation}
This manifold has a conifold singularity along the matter curve $C_m$
\begin{equation}
C_m: \qquad x\,\,=\,\,y=\,\,\psi\,\,=\,\,f\,\,=\,\,0\:.
\end{equation}
The most natural procedure to cure this singularity is to make a small resolution. This generates a new four-cycle $\hat{C}_m$, that is a $\mathbf{P}^1$-fibration over the curve $C_m$ (see also \cite{Esole:2011sm}).

The D3-tadpole contribution coming from the geometry of the 7-brane is given by
\begin{equation}
 Q^{\rm geom}_{D3} = -\tfrac{\chi(CY_4)}{24}\,.
\end{equation}
As the Euler number of the fourfold changes under the transition considered above (i.e. $g\equiv \psi^2-\rho$ $\rightarrow$ $g\equiv \psi^2$), the geometric contribution to the D3-tadpole changes as well. Since the total D3-charge must be conserved, 
an analogous change in the flux contribution must occur. This means that the $G_4$-flux studied before must undergo a transition into a new flux $\tilde{G}_4$. This new flux is constructed analogously as in \eqref{G4sm}, using the four-cycle
\begin{equation}\label{gamma4t}
\tilde{\gamma}_4:\qquad y=\psi \qquad\cap\qquad x=0 \qquad\cap\qquad \alpha=0 \:,
\end{equation}
where $\alpha$ is a polynomial of the base coordinates, whose degree is determined by requiring D3-charge conservation.

The induced chirality is given by the expression \cite{Donagi:2008ca, Hayashi:2008ba, Grimm:2011fx}.
\begin{equation}
I_{C_m}=\int_{\hat{C}_m} G_4^{I_1}\:.
\end{equation}

\

Let us follow this procedure in the weak coupling limit ($f=\epsilon\eta -3h^2$, $g=h(\epsilon \,\eta-2h^2) + \chi)$. 
Consider $\chi \equiv \psi^2$ (this corresponds to the `$U(1)$ restricted Tate model' \cite{Klemm:1996hh, Grimm:2010ez}).
In type IIB, the invariant D7-brane splits into a D7-brane and its image.
These branes intersect each other (away from the O-plane) along a curve $C_m$. 6D chiral modes live on this intersection. If there is a flux on the D7-brane, this will generate 4D chiral matter.
Correspondingly, the F-theory $CY_4$ develops a conifold singularity along curve $C_m$. 
As explained above, this is cured by a small resolution which produces a new four-cycle $\hat{C}_m$, which is a $\mathbf{P}^1$-fibration over $C_m$. The physical interpretation suggests that localized M2-branes wrapping the $\mathbf{P}^1$-fiber correspond to the 4D chiral matter.

The four-form flux $\tilde{G}_4$ can then be identified with the two-form flux on the D7-brane. This allows to compute and compare the number of chiral modes both in perturbative type IIB and directly in F-theory. The two numbers match: 
\begin{equation}
\tfrac12(\langle D7;D7\rangle  + \tfrac14
\langle O7;D7\rangle) = \int_{\hat{C}_m} G_4\:.
\end{equation}
The index on the l.h.s counts the chiral modes at the D7/image-D7 intersection (see \cite{Collinucci:2008pf}). 

Comparing the flux D3-tadpoles in type IIB and in F-theory, one obtains again a perfect match (see \cite{Braun:2011zm}).

\subsection{Non-abelian gauge groups}

Finally, we consider a more interesting setup: One (restricted) $I_1$-brane and one $Sp(1)$-stack. 
In the weak coupling limit this corresponds to a D7/image-D7 system and a stack of two D7-branes on an invariant divisor with an $Sp(1)$ gauge group.
In this situation, one has chiral modes at the  $Sp(1)$/$I_1$ intersection (besides the $I_1$-brane/image-brane chiral modes).
The F-theory $CY_4$ has one $Sp(1)$ singularity on the surface $S$ wrapped by the corresponding stack and a curve $C_m$ worth of conifold singularities. In addition, we also have $Sp(1)$-fundamental matter on the curve $C_f=S^{I_1}\cap S$ (where $S^{I_1}$ hosts the $I_1$ brane).

We cure the $Sp(1)$ singularity via a blow-up and the conifold one via a small resolution, see \cite{Collinucci:2010gz, Esole:2011sm} in this context. 
There will be a new divisor $E_{Sp}$ and two new four-cycles: $\hat{C}_f$, a $\mathbf{P}^1$ fibration over $C_f$; and $\hat{C}_m$, a $\mathbf{P}^1$ fibration over $C_m$. 

The $G_4$-flux, it is given by the sum of two objects
\begin{equation}
G_4=\tilde{G}_4^{(I_1)}+G_4^{(Sp)}
\end{equation}
where $\tilde{G}_4^{(I_1)}$ is like before (i.e. constructed using a four-cycle like \eqref{gamma4t}), and $G_4^{(Sp)}=\tfrac12 E_{Sp}\wedge p$ corresponds to the D7-brane flux along the Cartan generator of $Sp(1)$ ($p$ is some two-form living on $B_3$).

Comparing again with type IIB, the number of chiral modes on the various matter curves and the D3-tadpole all match. In particular 
\begin{equation}
\int_{C_f} F_{U(1)}-F_{Sp(1)} = \int_{\hat{C}_f} G_4 \:.
\end{equation}


\section{$G_4$-flux as a coherent sheaf}

In perturbative type IIB string theory, the D7-branes can be described in terms of coherent sheaves (see \cite{Collinucci:2008pf}). Physically, a D7-brane results from the partial annihilation between a D9- and an anti-D9-stack. 

Consider the exact sequence
\begin{equation}
 0 \rightarrow E_2 \xrightarrow[]{T} \tilde{E}_2 \rightarrow L_1|_{D7} \rightarrow 0 \qquad  \mbox{with} \qquad 
 T = \left(\begin{array}{cc}
 \xi\tau & -\eta-\xi\psi \\ \eta - \xi\psi & \xi\rho 
\end{array}  \right)
\end{equation}
$E_2$ and $\tilde E_2$ are rank two gauge bundles on anti-D9- and D9-stacks, respectively. T is the bifundamental tachyon matrix field with det$(T)=\eta^2 -\xi^2 (\psi^2-\rho\tau)$.

This set of data describes both the D7-brane and the flux living on it:
The divisor $S$ wrapped by the D7-brane is given by det$(T) = 0$ (D7 is a determinantal variety), while the flux is given by 
$F_2 = c_1(L_1)+\frac12 c_1(S)$. The beauty of this construction is that the geometry of the D7-branes and the fluxes are
unified in one object, which automatically takes care of their interplay. Furthermore, this formalism allows to deal with 
singularly shaped D7-branes and their fluxes by using sequences of vector bundles. This greatly facilitates calculations 
of the relevant indices.

\

Our idea is to generalize this formalism to F-theory. We first observe that the Calabi-Yau fourfold is not determinantal, it is instead Pfaffian. In fact, one can define an exact sequence in the ambient space $X_5$:
\begin{equation}
0\rightarrow E_4 \xrightarrow[]{M} \tilde{E}_4 \rightarrow V_2|_{CY_4} \rightarrow 0 
\qquad \mbox{where} \qquad
 M = \left(\begin{array}{cccc}
 0 & x & \rho & y+ \psi \\ -x & 0 & -y+\psi & \tau \\ 
 -\rho & y-\psi & 0 & x^2+f \\ -y-\psi & -\tau & -x^2-f & 0 \\ 
\end{array}   \right)\:.
\end{equation}
where $E_4$ and $\tilde E_4$ are now rank four vector bundles. $M$ is an antisymmetric matrix with Pfaff$(M) = (y + \psi ) (y - \psi) + \rho\tau - x (x + f )$.
Note the analogy with the tachyon matrix: The $CY_4$ hypersurface is given by Pfaff$(M)=0$, while the flux is given by $G_4 = c_2(V_2)-s$, 
where $s$ is some appropriately chosen four form in $H^2(X_4)\wedge H^2(X_4)$.
As shown in \cite{Braun:2011zm}, one can use this sequence to easily compute the D3-tadpole.

\section{Conclusions}

We have found an explicit realization of a large class of Poincar\'e invariant $G_4$-fluxes in F-theory, corresponding to a large class of $F_2$ fluxes on D7-branes. We have done this in terms of algebraic cycles. This flux is (by construction) globally defined and does not rely upon an extension of local fluxes. It restricts the complex struture of the fourfold in the same way as the $F_2$-fluxes restrict the corresponding D7-brane moduli.
The algebraic description of the flux makes it easy to compute phenomenologically important indices.
Whenever the type IIB weak coupling limit was available, we identified the constructed $G_4$-flux with $F_2$ fluxes on the involved D7-branes, finding a perfect
match in the D3-tadpole of the flux and the number of chiral modes it generates.

The fluxes considered here induce chirality, but do not break the non-abelian gauge groups. 
Similar results are contained in \cite{Grimm:2011tb,Krause:2011xj,Grimm:2011fx,Krause:2012yh}. 


Finally, we have given a description of $CY_4$'s together with $G_4$-fluxes as coherent sheaves. This is an exciting construction which should be explored further.

\section*{Acknowledgements}
We are grateful to Thomas Grimm, Arthur Hebecker, Christoph Mayrhofer and Timo Weigand for interesting and useful discussions.
The work of A. P. B. was supported by the FWF under grant I192. A. C. is a Research Associate of the Fonds de la Recherche 
Scientifique F.N.R.S. (Belgium). The work of R. V. was supported by the German Science Foundation (DFG) under the 
Collaborative Research Center (SFB) 676.

\end{document}